\documentclass[]{aastex631}
\usepackage{amssymb}
\usepackage{amsmath}
\usepackage{appendix}
\usepackage[T1]{fontenc}
\usepackage[utf8]{inputenc}

\shortauthors{Chen $\&$ Dai}
\shorttitle{Super-Eddington Disk Accretion onto NS}

\begin{document}
	
	\title{Super-Eddington Magnetized Neutron Star Accretion Flows: a Self-similar Analysis}
	
	\author[0000-0001-8955-0452]{Ken Chen}
	
	\affiliation{School of Astronomy and Space Science, Nanjing University, 
		Nanjing 210023, China}
	
	\author[0000-0002-7835-8585]{Zi-Gao Dai}
	
	\affiliation{Department of Astronomy, 
		University of Science and Technology of China, Hefei 230026, China; daizg@ustc.edu.cn}
	
\begin{abstract}

The properties of super-Eddington accretion disks exhibit substantial 
distinctions from the sub-Eddington ones. In this paper, we investigate
the accretion process of a magnetized neutron star (NS) surrounded by a
super-Eddington disk. By constructing self-similar solutions for the disk 
structure, we study in detail an interaction between the NS magnetosphere 
and the inner region of the disk, revealing that this interaction takes 
place within a thin boundary layer. The magnetosphere truncation radius 
is found to be approximately proportional to the Alfv\'{e}n radius, with 
a coefficient ranging between $0.34-0.71$, influenced by the advection
and twisting of a magnetic field, NS rotation, and radiation emitted from
an NS accretion column. Under super-Eddington accretion, the NS can readily
spin up to become a rapid rotator. The proposed model can be employed 
to explore the accretion and evolution of NSs in diverse astrophysical 
contexts, such as ultraluminous X-ray binaries or active galactic nucleus
disks.
	
\end{abstract}


\section{Introduction}
Accretion is a widespread process occurring in various neutron star 
(NS) systems, such as newborn NSs within supernova envelopes 
\citep[e.g.][]{Chevalier89}, NSs in high-mass X-ray binaries 
\citep[e.g.][]{Reig11}, and NSs in low-mass X-ray binaries 
\citep[e.g.][]{Bhattacharya91}, among others,
contributing significantly to the evolution of the mass, spin, and magnetic
field properties of the NSs \citep[e.g.][]{Konar17, Igoshev21, Abolmasov24},
while concurrently producing observable radiation 
\citep[e.g.][]{Mushtukov22, Ascenzi24}. 
Therefore, accretion plays a crucial role in the NS populations 
\citep{Harding13}. The accretion of the NS is distinctive. Due to the inherent 
magnetization of the NS, accreting matter is blocked by the NS 
magnetosphere from falling freely onto its solid surface, but can 
penetrate into the magnetosphere via instability and subsequently flows 
along the field lines toward the NS magnetic poles \citep[e.g.][]{Bozzo08, 
Shakura17, Fornasini23}. Additionally, a circum-NS disk can form 
when the accretion material possesses angular momentum, leading to a
slightly more complicated accretion process \citep[e.g.][]{Romanova15}.

So far, the majority of studies have primarily focused on 
sub-Eddington magnetized NS accretion systems, characterized by a
thin and approximately Keplerian rotating circum-NS disk. However,
there exist astrophysical scenarios that involve 
super-Eddington disk-accreting NSs. Recently,
the discovery of ultraluminous X-ray pulsars (ULXPs) with apparent 
luminosities ranging between $10^{39}-10^{41}\,\rm{erg}\,\rm{s}^{-1}$,
although with uncertain geometrical beaming \citep{Lasota23}, implies the 
existence of super-Eddington NS accretion systems \citep{Fabrika21, King23}.
Besides, NSs are predicted to widely reside in active galactic 
nucleus (AGN) disks \citep[e.g.][]{McKernan20, Tagawa21, Perna21}, where the
environment is extremely dense and differentially rotating \citep[e.g.][]{SG03}, 
thereby leading to a super-Eddington disk accretion of the NS 
\citep[e.g.][]{Pan21, Chen23}. In contrast to the standard thin accretion disk,
the super-Eddington disk exhibits distinctive properties, in that it is
geometrically thick, sub-Keplerian in rotation, and advection-dominated in
cooling \citep[e.g.][]{Abramowicz88}; additionally, the accretion process is 
always accompanied by a powerful outflow to reduce the inward inflow mass rate 
\citep{Begelman99, Gu15}. Therefore, it is imperative to conduct a 
comprehensive study of the super-Eddington magnetized NS accretion system,
enabling us to explore specific properties of NS accretion and the
evolution of the NS driven by a super-Eddington flow. Several works
have both analytically and numerically investigated the accretion structure 
of this system \citep[e.g.][]{Ohsuga07b, Takahashi17, Chashkina17, 
Chashkina19, Abarca21, Inoue23}.

\begin{figure*}
	\begin{center}
		\includegraphics[width=0.45\textwidth]{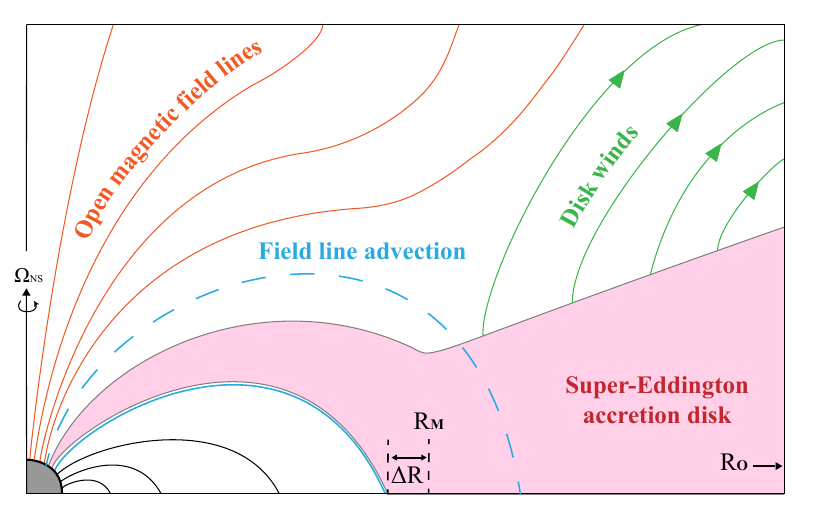}
	\end{center}
	\caption{
		Schematic diagram illustrating the super-Eddington disk accretion process 
		of a magnetized rotating NS, where the certain object is an NS,
		$\Omega_{\rm{NS}}$ is the NS spin rate, $R_{\rm{M}}$ denotes the 
		magnetosphere truncation radius, $\Delta R$ represents the thickness 
		of the interaction boundary layer, and $R_{\rm{o}}$ is the disk outer
		boundary, which is simplified set to be smaller than $R_{\rm{tr}}$, 
		indicating that the launching of disk winds occurs throughout
		the entire disk region.
		Some NS closed magnetic field lines, depicted by the blue dashed line,
		would undergo inward advection within the super-Eddington disk.
		Consequently, these field lines become concentrated at the
		magnetosphere-disk interaction layer, as indicated by the blue line,
		leading to an increase in magnetic pressure. It should be noted that
		the schematic serves solely for illustrative purposes and does not 
		accurately depict the relative scale of the system components,
		while also the interaction between the NS open field lines and the 
		disk winds is disregarded.}	
	\label{Fig:sketch}
\end{figure*}

In this paper, we investigate the super-Eddington accretion process
of a magnetized NS by constructing a generalized super-Eddington
circum-NS disk with varying strengths of the disk wind,
particularly focusing on an interaction between the NS magnetosphere 
and the inner disk. A schematic diagram of the accretion system is 
shown in Figure \ref{Fig:sketch}. This paper is organized as follows. 
In Section \ref{section2}, we present a self-similar solution to describe 
the structure of a super-Eddington disk. The specific 
magnetosphere-disk interaction process, as well as the resulting NS 
accretion properties, are shown in Section \ref{section3}. Several 
discussions are provided in Section \ref{section4}. Our conclusions
are summarized in Section \ref{section5}.
The terms $c$ and $G$ in this paper denote the speed of light and the 
gravitational constant, respectively. The mass of an NS is taken to be
$1.4M_{\odot}$ in the paper.

\section{Super-Eddington accretion disk}
\label{section2}
\subsection{Disk Equations}
In a super-Eddington accretion disk, internal photons are trapped 
because of a dense gas obstructing their vertical diffusion. A large 
amount of photons and energy can be advected inward with the gas 
inflow, rather than escape as radiation. When the timescale of 
photon vertical diffusion exceeds that of disk accretion, the 
radiation-effective flow converts to be radiation-ineffective roughly 
at the trapping radius \citep[e.g.][]{Kato08, Kitaki21}:
\begin{equation}
R_{\rm{tr}}=3 \dot{m} h R_{\rm{g}},
\end{equation} 
where $\dot{m}=\dot{M}/\dot{M}_{\rm{Edd}}$ is the dimensionless mass 
inflow rate of the disk, $\dot{M}_{\rm{Edd}}=L_{\rm{Edd}}/c^2$ is the 
Eddington limit accretion rate,  
$L_{\rm{Edd}}=4\pi G M m_pc/\sigma_T=1.26\times10^{38}m\,\rm{erg\,s}^{-1}$,
$m=M/M_\odot$ is the dimensionless mass of the NS, $m_p$ is the proton 
mass, $\sigma_T$ is the Thompson cross section, $h=H/R$ is the 
height ratio of the disk, and $R_{\rm{g}}= G M/ c^2$ is the
gravitational radius. Inside $R_{\rm{tr}}$, the trapped photons 
exert strong radiation pressure, which can induce disk winds to remove
mass, angular momentum, and energy from the accretion disk 
\citep{Ohsuga07a}. 

Containing the contribution of an outflow, the equation of continuity is 
written as
\begin{equation}
\frac{1}{R}\frac{d}{d R}\left(R\Sigma V_{\rm{R}}\right)+\frac{1}{2\pi R}\frac
{d \dot{M}_{\rm w}}{d R} = 0 \ , \label{EqM}
\end{equation}
where $\Sigma=2 \rho H$, $\rho$, and $V_{\rm{R}}$ are the surface 
density, density, and radial inflow velocity of the disk, respectively, 
and the outflow mass-loss rate $\dot{M}_{\rm{w}}$ is
\begin{equation}
\dot{M}_{\rm w}(R) = \int_{R_{\rm{in}}}^{R} 4\pi R^\prime\dot{m}_{\rm w}
(R^\prime)dR^\prime \ , \label{Mw}
\end{equation}
where $R_{\rm{in}}$ denotes the radius of the disk inner edge and
$\dot{m}_{\rm w}$ is the mass-loss rate per unit area from each disk face.

The integrated radial momentum equation and the azimuthal equation of motions,
respectively, can be written by 
\begin{equation}
V_{\rm R}\frac{dV_{\rm R}}{dR} + \left(\Omega_{\rm K}^2-\Omega^2\right)R + 
\frac{1}{\rho}\frac{dP}{dR} = 0 \ , \label{Eqr}
\end{equation}
\begin{equation}
-\frac{1}{R}\frac{d}{d R}\left(R^3\Sigma V_{\rm R}\Omega \right)+\frac{1}{R}
\frac{d}{d R}\left(R^3\nu \Sigma\frac{d \Omega}{d R}\right)-
\frac{\left(lR\right)^2\Omega}{2\pi R}\frac{d \dot{M_{\rm w}}}{d R} = 0 \ ,
\label{EqL}
\end{equation}
where $P$ and $\nu=-\alpha P/(\rho R d\Omega/dR)$ are the pressure and 
viscosity of the disk, $\alpha$ is the viscosity parameter \citep{SS73}, 
and $\Omega_{\rm K}$ is the Keplerian angular velocity. 
In the last term of Equation (\ref{EqL}), $l\geqslant1$ quantifies the 
specific angular momentum carried away by the disk wind,
where $l=1$ represents the outflow removing the local disk 
angular momentum at its originating radius
\citep[e.g.][]{Knigge99, Proga00}.

The energy equation is 
\begin{equation}
Q_{\rm{vis}} = Q_{\rm adv} + Q_{\rm rad} + Q_{\rm w} \ , \label{energy}
\end{equation}
where $ Q_{\rm{vis}}$, $Q_{\rm adv}$, and $ Q_{\rm rad}$ are the viscous
heating rate, the advection cooling rate, and the radiation cooling rate
per unit area, respectively, for which the expressions are as follows 
\citep[e.g.][]{Abramowicz95, SG03}:
\begin{equation}
Q_{\rm{vis}} = \nu \Sigma\left(R\frac{d\Omega}{dR}\right)^2 \ , \label{Qvis}
\end{equation}
\begin{equation}
Q_{\rm adv} = \Sigma V_{\rm R} T\frac{dS}{dR} = \Sigma V_{\rm R}\left(\frac{1}
{\gamma-1}\frac{dc_{\rm s}^2}{dR} - \frac{c_{\rm s}^2}{\rho}\frac{d\rho}{dR}
\right) \ , \label{Qadv}
\end{equation}
\begin{equation}
Q_{\rm rad} = \frac{16}{3}\sigma T^4
\left(\tau + \frac{4}{3}+\frac{2}{3\tau}\right)^{-1} \ , \label{Qrad}
\end{equation}
where $S$ is the specific entropy, $\gamma$ is the ratio of specific 
heat, $c_{\rm s}=(P/\rho)^{1/2}$ is the sound speed, and $\tau=\kappa \rho H$ 
is the optical depth, for which the electron scattering is included 
as the main opacity source, $\kappa=\kappa_{\rm{es}}=0.4$.
The quantity $Q_{\rm w}$ in Equation~(\ref{energy}) represents the energy
carried away by the outflow, which can be expressed by
\begin{equation}
Q_{\rm w} = 2\eta \left(\frac{1}{2}\dot{m}_{\rm w} V_{\rm K}^2\right) \ , \label{Qw}
\end{equation}
where a factor of 2 signifies the outflow that is emitted bilaterally 
from the accretion disk, $\eta$ is a specific energy parameter
to determine the outflow strength, and $V_{\rm K}$ is the 
Keplerian velocity.

Due to the presence of the disk wind, the mass inflow rate would reduce 
inward. Numerical simulations show that the reduction roughly follows
a power law within $R_{\rm{tr}}$ 
\citep[e.g.][]{Yang14, Kitaki21, Yoshioka22, Yang23},
so we take the mass inflow rate varying with radius as \citep{Begelman99}:
\begin{equation}
\dot{M} =-2\pi R\Sigma V_{\rm{R}}=\dot{M}_{\rm{o}}\left(\frac{R}
{R_{\rm{o}}}\right)^{s} \ , \label{eqacc}
\end{equation}
for a disk with $R_{\rm{o}}<R_{\rm{tr}}$, where $\dot{M}_{\rm{o}}$ 
is the mass accretion rate at the disk outer boundary $R_{\rm{o}}$, 
and the power-law index $s$ is a free parameter, constrained within
$0 \leqslant s \leqslant 1$, which approaches zero in the absence of wind, 
leading to the disk satisfying the slim solution 
\citep[e.g.][]{Abramowicz88, Narayan95, Watarai99}.
Assuming that the complete reduction 
of mass inflow rate is solely attributed to 
the disk wind, then, from Equation (\ref{EqM}) and (\ref{eqacc}),
we can get
\begin{equation}
\dot{m}_{\rm w}= \frac{s \dot{M}}{4 \pi R^2}.
\end{equation} 

\subsection{Self-similar solutions}
Although numerical simulations are the more reliable approach for 
investigating the complicated super-Eddington accretion 
process around compact objects 
\citep[e.g.][]{Ohsuga05, Jiang14, Sadowski15, Kitaki18, Takahashi18, 
Abarca18, Hu22}, analytic solutions are valuable, as
they can roughly match the structure of the accretion systems 
observed in simulations \citep[e.g.][]{Jiao15}. Moreover, these solutions 
offer convenience in directly constructing circum-CO disks under 
varying system parameters
\citep[e.g.][]{Begelman99, Begelman12, Gu15, Ghoreyshi20, Zahra20}. 
Because a super-Eddington accretion disk is radiation-ineffective, 
the radiation cooling term is of minor significance 
in Equation (\ref{energy}) \citep[e.g.][]{Wu22}. If we ignore 
$Q_{\rm rad}$, a set of self-similar variables\textendash
$\rho_{\rm{s}}\varpropto R^{s-3/2}$, $P_{\rm s}\varpropto R^{s-5/2}$, 
$V_{\rm{R,s}}\varpropto R^{-1/2}$, $H_{\rm{s}}\varpropto R$, 
and $\Omega_{\rm{s}}\varpropto r^{-3/2}$\textendash can simultaneously satisfy 
the disk Equations (\ref{EqM}), (\ref{Eqr}), (\ref{EqL}), and (\ref{energy}).
Considering the vertical force balance, the disk height can be 
expressed as $H=c_{\rm s}/\Omega_{\rm K}$ \citep[e.g.][]{Kato08}. 
By combining these five equations, a solution to the disk 
structure can be obtained.

\begin{figure*}
	\begin{center}
		\includegraphics[width=0.415\textwidth]{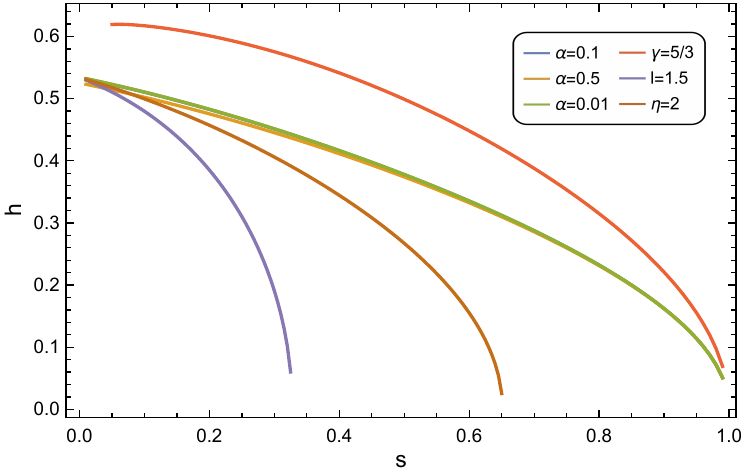}
		\includegraphics[width=0.46\textwidth]{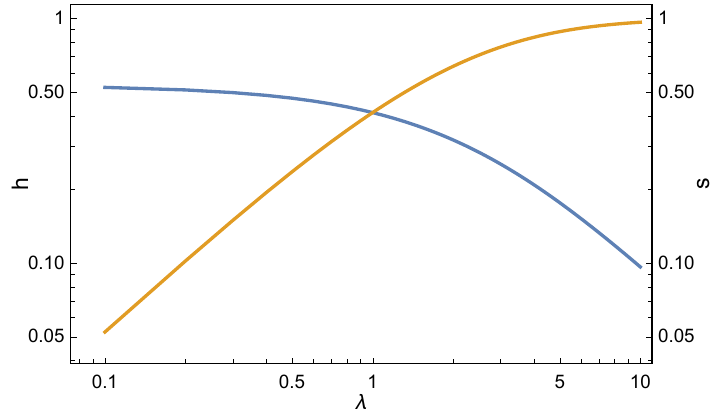}
	\end{center}
	\caption{Properties of a super-Eddington accretion disk under 
		self-similar solutions. The left panel represents the disk thickness 
		$h=H/R$ as a function of the inflow power-law index $s$ considering various
		system parameters (as shown in the inset), where the fiducial ones are
		$\alpha=0.1$, $\gamma=4/3$, $l=1$, and $\eta=1$. The right panel shows the 
		variation of $h$ and $s$ as a function of $\lambda$, where 
		$\lambda=s/(H/R)$ qualitatively reflects the impact of advection on 
		the disk. }	
	\label{Fig:sh}
\end{figure*}

The self-similar solution is:
\begin{equation}\label{rhos}
	\rho_{\rm{s}} = \frac{ \left(1+2s-2l^2s\right) \dot{M}_{\rm o} \omega}
	{4 \pi \left(1+2s\right) \alpha (G M)^{1/2} R_{\rm o}^{3/2} h^3} \left(\frac{R}{R_{\rm o}}\right)^{s-3/2},
\end{equation}
\begin{equation}\label{Ps}
	P_{\rm{s}} = \frac{\left(1+2s-2l^2s\right) \dot{M}_{\rm o} (G M)^{1/2} \omega}
	{4 \pi \left(1+2s\right) \alpha R_{\rm o}^{5/2} h } \left(\frac{R}{R_{\rm o}}\right)^{s-5/2},
\end{equation}
\begin{equation}\label{Vrs}
	V_{\rm{R,s}} = - \frac{ \left(1+2s\right) \alpha (G M)^{1/2} h^2}
	{\left(1+2s-2l^2s\right) \omega R_{\rm o}^{1/2}} \left(\frac{R}{R_{\rm o}}\right)^{-1/2},
\end{equation}
\begin{equation}\label{Hs}
    H_{\rm{s}} = h R,
\end{equation}
\begin{equation}\label{omegas}
	\Omega_{\rm{s}} = \omega  \frac{(G M)^{1/2}}{R_{\rm o}^{3/2}} \left(\frac{R}{R_{\rm o}}\right)^{-3/2},
\end{equation}
where
\begin{equation}
	\omega = \frac{1}{2} \left\{ \left(2s-5\right) h^2 +
	\left[ \left( \left(5-2s\right)^2-
	\frac{8 \alpha\left(1+2s^2\right)}{1+2s-2l^2s}\right) h^4
	+4\left(2s-5\right) h^2 + 4\right]^{1/2}
	+ 2\right\}^{1/2} ,
\end{equation}
and $h$ is solved numerically via Equation (\ref{energy}), i.e.,
\begin{equation}\label{h}
    \frac{3\left(1+2s-2l^2s\right)}{4\left(1+2s\right)}
    \left\{ \left(2s-5\right)h^2+
    \left[\left(2+\left(2s-5\right)h^2\right)^2-
    \frac{8\left(2s+1\right)^2\alpha^2h^4}{\left(1+2s-2l^2s\right)^2}
    \right]^{1/2}+2\right\}
    + \left(3-2s-\frac{2}{\gamma-1}\right)h^2
    -\eta s=0. 
\end{equation}

As indicated by Equations (\ref{rhos})-(\ref{omegas}), the disk properties 
are influenced by $s$, $\alpha$, $\gamma$, $l$, and $\eta$. To investigate
these dependencies, we take $h$ as an indicator and explore its variation, 
which is shown in the left panel of Figure \ref{Fig:sh}. 
First, the super-Eddington accretion disk is geometrically thick, 
with $h>0.1$. The value of $h$ decreases with the increase of $s$,
which can be attributed to the enhanced mass loss resulting from a larger
$s$. Consequently, more energy is carried away by the disk wind, leading 
to a reduction in the thickness of the resultant disk. 
Second, for a large value of $l$ or $\eta$, the maximum value of $s$ in 
a valid solution is less than $1$, since a disk would not exist if the 
loss of angular momentum or energy via the disk wind were to
exceed its total reserve. Third, the disk structures are basically not
influenced by $\alpha$, but exhibit sensitivity to $\gamma$. 
In the subsequent calculations, we set $\alpha=0.1$, $\gamma=4/3$, $l=1$, 
and $\eta=1$. 

Moreover, the selection of $s$ should be based on the disk properties 
rather than being chosen randomly. \cite{Wu22} adopt a reasonably appropriate 
assumption that $s$ is proportional to $H/R$, i.e., $s=\lambda(H/R)$, 
where $\lambda$ is a constant, considering that the disk thickness can reflect 
the strength of advection. Still, $\lambda$ is uncertain, but it can be adjusted 
to encompass all values of $s$, as shown in the right panel of Figure \ref{Fig:sh}.
Therefore, we apply all physical values for $s$ from $0$ to $1$ to investigate
the comprehensive properties of the super-Eddington accretion disk.

\section{Super-Eddington disk accretion of magnetized NS}
\label{section3}
The fundamental framework of disk accretion around a magnetized NS is widely 
accepted in theory works (e.g. \citealt{Pringle72, Lamb73}, 
and see \citealt{Lai14} for a review) and supported by numerical simulations 
\citep[e.g.][]{Romanova04, Romanova12, Zhu24}.
According to this framework, the circum-NS disk is disrupted 
by the NS magnetic field due to the high pressure 
and stress it exerts at the magnetosphere truncation radius. Subsequently, 
within an interaction boundary layer, the angular velocity of the inflow  
transitions from the disk value to the NS spin rate through magnetic torque. 
The gas, coupled with the rotating NS, is then loaded onto closed field lines 
and funneled to the magnetic poles. Using the disk structures constructed in 
the previous section, we investigate the specific properties of super-Eddington
accretion around the magnetized NS. 

\subsection{Magnetosphere Truncation Radius}
The external magnetic field of an NS exhibits a complex structure
superposed by various multipoles \citep[e.g.][]{Bilous19}. Since the $n-$th 
multipole field varies as $B_{n}\sim\mu_{n}/R^{n+2}$, where $\mu_{n}$ is 
the $n-$th magnetic moment, in regions far 
from the NS surface, higher multipoles fall off rapidly and the dipole 
dominates the overall field structure\footnote{The dipole magnetosphere
truncation radius $R_{\rm{M}}= 3.9\times 10^7 \,\rm{cm} \, k_{\rm{in}} 
\left(\mu/10^{30}\,\rm{G}\,\rm{cm}^3\right)^{4/7}
\left(M/1.4 M_\odot\right)^{-1/7}
\left(\dot{M}_{\rm{in}}/10^{3}\dot{M}_{\rm{Edd}}\right)^{-2/7}$ is much larger
than the NS radius, so the multipoles are dominant only if 
$\mu_{n}\gg \mu_{\rm{dipole}}$. However, near the NS surface, the multipole 
fields can be comparable or stronger than the dipole component. So, the magnetic 
field at the NS surface exhibits a complex structure due to the superposition of 
multipoles with varying tilt angles, guiding the funneled inflow to a area 
wider than the pure dipole magnetic poles \citep[e.g.][]{Long07, Long08, Gregory11}.
Consequently, the accretion streams would overlap and heat various surface regions 
of the NS, thereby generating anomalous multiwavelength radiation facilitated by 
the NS rotation \citep[e.g.][]{Bilous19}. Additionally, in the case of super-Eddington 
accretion, the funneled streams can form an optically thick envelope that can
then reprocess the photons emitted by the multipole accretion flows encircled in 
the closed dipole fields, resulting in a more intricate radiation profile
\citep{Mushtukov19}.}. Therefore, at the 
magnetosphere truncation radius, we adopt a dipole structure for the NS 
magnetic field with a field strength of $B_0=\mu/R^3$, where 
$\mu= 10^{30}\,\rm{G\,cm^3}$ is set. Simulations indicate that
the formation of a funnel stream occurs approximately at the radius where 
magnetic and disk pressure reach an equilibrium \citep{Kulkarni13, Takahashi17}. As the 
inflow mass rate of a super-Eddington disk exhibits radial variation, 
we employ pressure balance as the criterion for determining the magnetosphere 
truncation radius $R_{\rm{M}}$ \footnote{The ram pressure of the accreting 
gas is negligible compared to the thermal pressure of the disk, as
$\rho_{\rm{s}} V_{\rm{R,s}}^2/P_{\rm{s}}=\left[ \left(1+2s \right)\alpha h/
\left(1+2s-2l^2 s\right) \omega \right]^2$ is much less than $1$, except in 
the case where $\alpha \sim 1$ with $s \sim 0$. Therefore, we disregard the 
contribution of this term to the pressure balance at $R_{\rm{M}}$.}, 
at which the disk gas becomes trapped by 
field lines and the mass rate accreted onto the NS is then 
$\dot{M}_{\rm{o}}(R_{\rm{M}}/R_{\rm{o}})^{s}$, i.e.:
\begin{equation}
	P_{\rm{B0}}=P_{\rm{s}}, \ \ \ 
	\frac{\mu^2}{8 \pi R_{\rm{M}}^6}
	=\frac{\left(1+2s-2l^2s\right) \dot{M}_{\rm o} (G M)^{1/2} \omega}
	{4 \pi \left(1+2s\right) \alpha R_{\rm o}^{5/2} h } \left(\frac{R}{R_{\rm o}}\right)^{s-5/2} ,
\end{equation}
which leads to
\begin{equation}
	R_{\rm{M}}=k\left(\frac{\mu^4 R_{\rm{o}}^{2s}}{G M \dot{M}_{\rm{o}}^2}\right)^{1/(7+2s)} ,
\end{equation}
where
\begin{equation}
	k = \left[\frac{(1+2s)\alpha h}{2(1+2s-2l^2s)\omega}\right]^{2/(7+2s)}
	\label{eqk}
\end{equation}
describes the dependence of the magnetosphere truncation radius on the properties of
a super-Eddington accretion disk. More precisely, the truncation is directly 
determined by the interaction between the NS magnetosphere and the innermost region 
of the disk, therefore we also present $R_{\rm{M}}$ in a widely accepted form 
\citep[e.g.][]{Ghosh79a, Ghosh79b}:
\begin{equation}
	R_{\rm{M}}=k_{\rm{in}}\left(\frac{\mu^4}{G M \dot{M}_{\rm{in}}^2}\right)^{1/7} ,
\end{equation}
where the second factor on the right-hand side is the conventional expression for 
the magnetosphere radius in the NS spherical accretion system \citep[e.g.][]{Elsner77}, 
i.e., the Alfv\'{e}n radius, with $\dot{M}_{\rm{in}}=\dot{M}(R_{\rm{M}})$ and
\begin{equation}
	k_{\rm{in}}= \left[\frac{(1+2s)\alpha h}{2(1+2s-2l^2s)\omega}\right]^{2/7}.
	\label{eqkin}
\end{equation}

\begin{figure*}
	\begin{center}
		\includegraphics[width=0.45\textwidth]{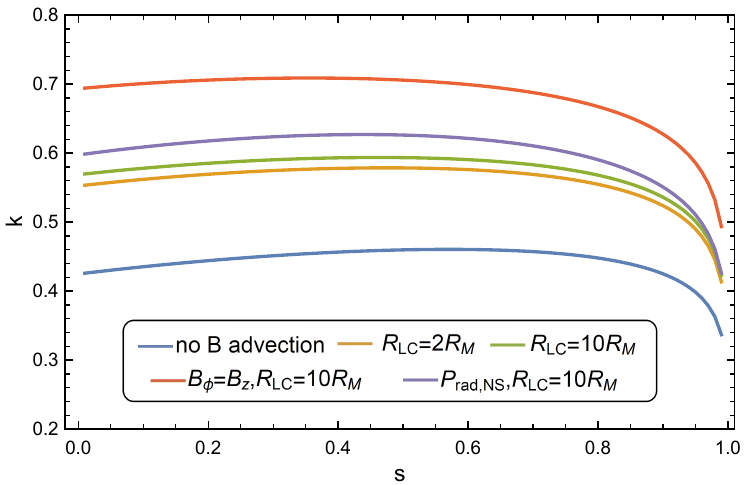}
		\includegraphics[width=0.45\textwidth]{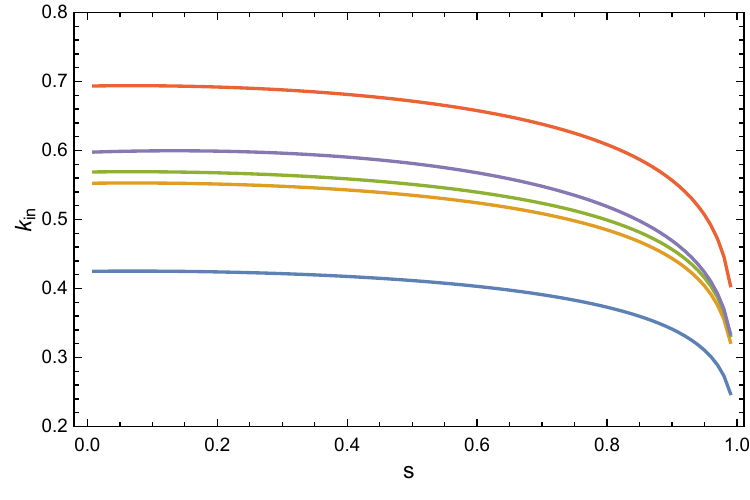}
	\end{center}
	\caption{Variation of magnetosphere truncation radius as a function of 
		the super-Eddington accretion disk's index $s$, where $k$ and 
		$k_{\rm{in}}$ are given by Equation (\ref{eqk}) and (\ref{eqkin}), 
		respectively. In each panel, the blue line shows the closed 
	    NS magnetic field lines initially threading the disk are 
	    inefficiently advected inward with the inflow, while the disk 
	    truncation results from the local pressure balance, i.e.,
	    $P_{\rm{B0}}=P_{\rm{s}}$; the golden line shows these closed  
        lines are concentrated at the magnetospheric boundary to enhance the 
        magnetic field strength, with the NS rotating at an angular velocity
        with $R_{\rm{LC}}=2R_{\rm{M}}$, while the green line correspondingly
        shows the case of a rotating NS with $R_{\rm{LC}}=10R_{\rm{M}}$; 
        the orange line shows the compressed dipole lines are strongly
        twisted to produce a toroidal component with $B_\phi=B_{\rm{z}}$, which
        provides extra magnetic pressure; and the purple line considers the extra
        feedback of the radiation released near the NS on the interaction
        boundary layer. }	
	\label{Fig:k}
\end{figure*}

To investigate the impact of the super-Eddington flow on the magnetized NS accretion, 
we examine the correlation between $k$ or $k_{\rm{in}}$ and $s$ to assess if it 
deviates significantly from the typical value $k_{\rm{in}}\sim0.4-0.8$ 
\citep[e.g.][]{Long05, Bessolaz08, Kulkarni13, Zanni13}. As shown in Figure \ref{Fig:k},
the value of $k$ ranges from $0.34$ to $0.46$, which is similar but relatively 
low compared to the typical value, owing to the high pressure of the
super-Eddington disk resulting from significant energy storage and advection 
processes. Furthermore, the variation trend of $k$ on $s$ shows an initial increase 
followed by a subsequent decrease, which mirrors the behavior of the disk pressure 
(not depicted in the paper, for conciseness). Notably, $k$ exhibits minimal variation 
with the change in $s$, except for $s>0.9$, where a substantial compression
of the NS magnetosphere caused by high $P_{\rm{s}}$ leads to a reduction 
of $k$. Correspondingly, $k_{\rm{in}}$ varies with the range of $0.25-0.43$, 
which is slightly lower than $k$. Moreover, taking $\dot{M}_{\rm{in}}$ into account, $k_{\rm{in}}$ exhibits a monotonic decrease as the value of $s$ increases.

When the NS closed magnetic field lines thread the accretion disk, 
they simultaneously undergo inward advection, along with inflow 
and outward diffusion controlled by turbulence \citep[e.g.][]{Parfrey17}. 
The advection of the magnetic field relies on 
the disk radial velocity $V_{\rm{R,s}}$, i.e., Equation (\ref{Vrs}), 
while its diffusion is determined by the magnetic diffusivity $\eta$.
In magnetorotational-instability-generated turbulence, $\eta$ can
be expressed as $\eta=\nu/Pr_{\rm{m}}$, 
where $\nu=2\alpha c_{\rm{s}} H/3\omega$
and the magnetic Prandtl number $Pr_{\rm{m}} \sim 1$
\citep[e.g.][]{Lesur09, Guan09, Fromang09}, 
while the resulting diffusion velocity 
can be approximated as $V_{\rm{diff}}\sim \eta/H$
\citep[e.g.][]{Lubow94, Cao11}. 
For a super-Eddington accretion disk, using a self-similar solution,
we find that $V_{\rm{R,s}}\gtrsim V_{\rm{diff}}$, owing to an 
effective disk advection. Additionally, as the wind 
extracts extra angular momentum from the disk, it would lead to 
$Pr_{\rm{m}}$ exceeding unity \citep{Parfrey17}. Consequently, 
the threading magnetic field is expected to overcome diffusion and flow inside.
To estimate the impact of the field advection, we consider an ideal scenario 
where all closed magnetic field lines originating from the NS and initially 
penetrating the accretion disk are advected inward and concentrated at the 
magnetosphere-disk interaction layer.
Adopting the magnetic flux conservation
\begin{equation}
\delta\left(B \times Area\right)=0, \ \ \ 
\delta B \pi R^2+ \frac{\mu}{R^3}\left(-2\pi R \delta R\right)=0,
\end{equation}
the compressed magnetic fields can be derived as
\begin{equation}
	\Delta B = \int^{R_{\rm{LC}}}_R \delta B =
	\int^{R_{\rm{LC}}}_R \frac{2\mu}{R^4} d R =\frac{2}{3}\mu
	\left(\frac{1}{R^3}-\frac{1}{R_{\rm{LC}}^3}\right),
\end{equation}
where $R_{\rm{LC}}$ is the light cylinder radius, inside which the initial NS
magnetic field lines are closed.
Therefore, the magnetic field at the magnetosphere truncation radius 
can be estimated by superimposing the compressed component onto the initial 
value, i.e.:
\begin{equation}
    B=B_0+\Delta B=\frac{\mu}{R^3}
    \left(\frac{5}{3}-\frac{2}{3}\frac{R^3}{R_{\rm{LC}}^3}\right).
    \label{B}
\end{equation}

Due to the concentration of the magnetic field, a higher magnetic pressure, 
$P_{\rm{B}}=B^2/8 \pi$, is generated, which pushes the disk outward,
consequently resulting in an enlarged magnetosphere truncation radius. 
Two instances of a rotating NS with 
$R_{\rm{LC}}=2R_{\rm{M}}$ and $R_{\rm{LC}}=10R_{\rm{M}}$ are illustrated 
in Figure \ref{Fig:k}, the corresponding values of $k$ are raised to $0.41-0.58$ and
$0.42-0.59$, respectively. The similarity between the two values suggests 
that $k$ or $k_{\rm{in}}$ is not significantly affected by the NS rotation, 
which is consistent with Equation (\ref{B}), where the second term, 
$\propto R_{\rm{LC}}^{-3}$, remains negligible unless the NS rotates very fast, 
so that $R_{\rm{LC}}\sim R_{\rm{M}}$. The compressed magnetic field 
asymptotically approaches $B\simeq 5/3B_{0}$.

Besides advection and diffusion, the field lines linking the NS and the disk are 
twisted due to the differential rotation at each end of the footpoints, resulting
in the generation of a toroidal field component. The resultant magnetic pressure 
drives field inflation and reconnection, opening the closed lines
to sever the NS-disk linkage, ultimately arresting the growth of the toroidal field
\citep{Aly90, Lovelace95, Parfrey16}. \cite{Lai14} argues that the maximum toroidal 
twist is $\lvert B_{\phi}/B_{z}\lvert_{\rm{max}}\sim 1$. By assuming 
$B_{\phi}=B_{z}$, the total magnetic pressure at the magnetosphere-disk 
interaction layer increases to $B^2/4\pi$, leading to a larger truncation radius.
As shown in Figure \ref{Fig:k}, the resulting values of $k$ and $k_{\rm{in}}$ are
$0.49-0.71$ and $0.40-0.69$, respectively, which are consistent with the typical
values and agree with the simulation results \citep[e.g.][]{Takahashi17}, wherein 
$k_{\rm{in}} \sim 0.7$. However, our findings slightly deviate from those of 
\cite{Chashkina17, Chashkina19}, where the relative magnetosphere radius, 
i.e., $k$ and $k_{\rm{in}}$, exhibits a significant dependence 
on the inner disk mass inflow rate. We attribute this discrepancy primarily to 
the differences in the boundary conditions employed.
It is worth noting that despite the high efficiency of advection, not all closed magnetic
field lines initially threading the accretion disk are swept inward because 
of magnetic inflation, flux opening, and reconnection \citep{Parfrey17}. 
The disk winds also exert a notable impact on the magnetic field structure. Moreover, 
reconnection prevents the twisted toroidal magnetic field from always remaining at its 
maximum strength. Therefore, the actual values of $k$ and $k_{\rm{in}}$ 
should be lower than those derived from the ideal conditions.

Additionally, when the super-Eddington accretion flow is funneled close to 
the NS magnetic pole, it gives rise to the formation of an accretion column 
\citep{Mushtukov15a}, within which the released gas energy is converted to 
radiation and subsequently escapes from its lateral surface 
\citep[e.g.][]{Mushtukov15b, Kawashima16, Mushtukov18}. When these photons from the 
NS illuminate the interaction boundary, an extra radiation pressure 
is exerted upon it \citep[e.g.][]{Chashkina19}. Considering this pressure component,
the equilibrium criterion is modified to
\begin{equation}
P_{\rm{B}}+P_{\rm{rad,NS}}=P_{\rm{s}}, \ \ \ 
P_{\rm{rad,NS}}=\frac{L_{\rm{rad,NS}}}{4\pi R_{\rm{M}}^2 c},
\end{equation}
where $L_{\rm{rad,NS}} \simeq G M \dot{M}_{\rm{in}}/R_{\rm{NS}}$ is the luminosity of 
the radiation emitted from the accretion column, $R_{\rm{NS}}=10^6\,\rm{cm}$ 
represents the NS radius, and the radiation is simplified to be isotropic. 
The value of $k$ and $k_{\rm{in}}$ for a case of contrast, as shown in 
Figure \ref{Fig:k}, ranges between $0.43-0.63$ and $0.34-0.60$, respectively. 
These values are slightly higher than those obtained when disregarding the 
contribution of the accretion column radiation, yet the deviation remains within $5\%$,
suggesting that the impact of this radiation effect on disk truncation 
is insignificant.

In brief, the magnetosphere truncation radius, which is influenced by the properties
of the super-Eddington disk and determines 
the mass accretion rate onto the NS, is found to be
proportional to the Alfv\'{e}n radius, as described by Equations (\ref{eqk}) 
and (\ref{eqkin}), where $k$ and $k_{\rm{in}}$ vary within the ranges of 
$0.34-0.71$ and $0.25-0.69$, respectively, depending on various effects, 
including the advection and twisting of the magnetic field, 
the rotation of the NS, as well as the
radiation emitted from the NS accretion column.
The resulting mass rate accreted onto the NS is approximated as
$\dot{M}_{\rm{in}}= \dot{M}_{\rm{o}}(R_{\rm{M}}/R_{\rm{o}})^s$.

\subsection{Thickness of Interaction Boundary Layer}

\begin{figure*}
	\begin{center}
		\includegraphics[width=0.45\textwidth]{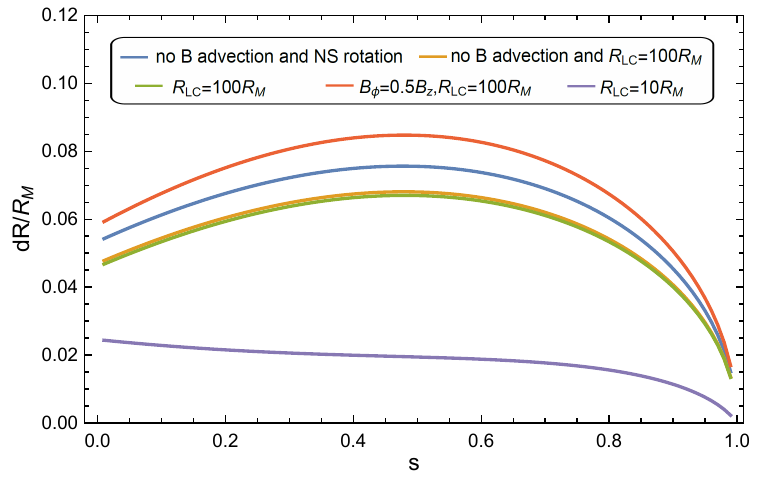}
	\end{center}
	\caption{Relative thickness of the magnetosphere-disk interaction 
		boundary layer, i.e., $\Delta R/ R_{\rm{M}}$, as a function of
		$s$ for a sample system with $\dot{M}_{\rm{o}}=10^4\dot{M}_{\rm{Edd}}$
		and $R_{\rm{o}}=10^4R_{\rm{g}}$. 
		The blue line represents the case of no magnetic field 
	    advection within the disk and the NS without spinning, i.e.,
        $R_{\rm{LC}}\rightarrow\infty$, while the golden line correspondingly
        shows the case of $R_{\rm{LC}}=100R_{\rm{M}}$; the green line 
        shows that the magnetic fields are concentrated and
        $R_{\rm{LC}}=100R_{\rm{M}}$, the orange line reduces the degree
        of field twisting to $B_\phi=0.5B_z$; 
        and the purple line speeds up the NS 
        rotation to $R_{\rm{LC}}=10R_{\rm{M}}$. $\Delta R/ R_{\rm{M}}\ll 1$
        always holds, indicating that the boundary layer is very thin. }	
	\label{Fig:dr}
\end{figure*}

At the truncation radius, though the NS magnetic pressure hinders the disk
inflow, various instabilities can arise to facilitate further plasma 
penetration into the magnetosphere (\citealt{Romanova15} and references therein).
The differential rotation between the NS and the magnetosphere plasma causes 
a twisting of the field lines, in turn resulting in the generation of 
magnetic torque to eliminate the rotational difference within a boundary layer 
\citep[e.g.][]{Spruit93, Rappaport04, Lai14}. 
In the case of a typical Keplerian disk, 
the interaction boundary layer is thin \citep[e.g.][]{Kulkarni13, Cemelji23};
for comparative analysis, we explore the thickness of this layer under
super-Eddington disk conditions. Taking into account 
the contribution of both the twisting magnetic field lines above and 
below the disk midplane, the magnetic torque per unit area 
can be expressed as $R_{\rm{M}} B_\phi B_z/2\pi$, facilitating the
transition of the inflow angular velocity from the disk's 
$\Omega_{\rm{s}}$ to the NS spin rate $\Omega_{\rm{NS}}$ 
within the boundary layer, i.e.:
\begin{equation}
    2\pi R_{\rm{M}}^2 \Delta R \frac{B_\phi B_z}{2\pi} 
    = R_{\rm{M}}^2 \Delta R B_\phi B_z
    =\dot{M}_{\rm{in}}\left| R_{\rm{M}}^2 \Omega_{\rm{s}}-
    (R_{\rm{M}}-\Delta R)^2 \Omega_{\rm{NS}} \right|,
    \label{dr}
\end{equation}
where $\Delta R$ is the layer radial thickness, and the absolute 
value quantifies the overall change in the inflow's specific angular momentum.

The relative thickness of the interaction boundary layer, 
$\Delta R/ R_{\rm{M}}$, under various conditions is shown in Figure
\ref{Fig:dr}; in all cases, unless otherwise specified, 
we assume that the field twisting 
results in $B_{\phi}=B_{z}$. When the NS spin rate is low, 
or even when the NS is approximately nonrotating, i.e., 
$\Omega_{\rm{NS}}\sim 0$, the second term in the absolute sign of 
Equation (\ref{dr}) can be neglected, and the layer thickness is reduced to
\begin{equation}
	\Delta R = \frac{\dot{M}_{\rm{in}} \Omega_{\rm{s}} R_{\rm{M}}^6}{\mu^2},
\end{equation}
which is thin, with a relative thickness smaller than $0.1$. The
variation trend of $\Delta R/ R_{\rm{M}}$ on $s$ 
exhibits a similar pattern to that of $R_{\rm{M}}$ or $k$, initially increasing 
and then decreasing, attributed to the layer thickness being determined
by the torque exerted by the magnetic fields, which is directly influenced by the 
magnetosphere truncation radius through $B \propto R_{\rm{M}}^{-3}$.
Spinning up the NS, for example to $R_{\rm{LC}}=100R_{\rm{M}}$, diminishes 
the gap in angular velocity between it and the disk, causing a thinner layer;
as a comparison, for an NS with $R_{\rm{LC}}=10R_{\rm{M}}$, its angular 
velocity closely matches that of the disk, leading to an extremely 
thin boundary.
The inclusion of field advection and accumulation leads to a decrease 
in the relative layer thickness, as a stronger field generates a larger 
torque; however, this effect is not highly significant. On the contrary,
when the degree of field twisting is low, for instance,
$B_\phi=0.5B_z$, the resulting magnetic torque is attenuated;
to fully transform the rotation of disk gas, a thicker
layer is required, for which the relative thickness still remains below $0.1$.
Overall, the interaction boundary layer between the NS magnetosphere
and the super-Eddington disk is thin, in turn ensuring the availability
of Equation (\ref{dr}), as both the NS magnetic field and the mass 
inflow rate remain relatively constant within a narrow $\Delta R$.

\section{Discussion}
\label{section4}
\subsection{Propeller regime}
In an NS disk accretion system, when the NS rotates slower than the disk
at its inner boundary, after being coupled with the magnetosphere, 
the braked inflow can no longer counteract gravity and becomes 
channeled toward the NS magnetic poles along the field lines. 
Conversely, when the NS is the faster object, the disk gas is accelerated 
by the magnetic torque to gain angular momentum, and the resulting enhanced 
centrifugal force hinders the subsequent accretion and may even 
lead to the ejection of most of the mass at the magnetosphere boundary,
forming an outflow; this process is called the propeller regime 
\citep[e.g.][]{Illarionov75, Lovelace99, Ustyugova06}. By defining a fastness 
parameter, $\omega_{s} = \Omega_{\rm{NS}} / \Omega_{\rm{s}}(R_{\rm{M}})$,
the NS is spun up through gas accretion when $\omega_{s}<1$ and is spun 
down by interaction torque when $\omega_{s}>1$. Consequently, the NS 
reaches an equilibrium state approximately at $\omega_{s} \simeq 1$ 
\citep[e.g.][]{Long05, Dai06, Lai14} and the corresponding NS spin period is
\begin{equation}\label{Peq}
	P_{\rm{eq}}=\frac{2 \pi }{\Omega_{\rm{s}}(R_{\rm{M}})}= 2 \pi 
	\omega^{-1} k_{\rm{in}}^{3/2} \left(G M\right)^{-5/7} 
	\left(\frac{\mu^2}{\dot{M}_{\rm{in}}}\right)^{3/7}
	= 0.03\,\rm{s} \left(\frac{\mu}{10^{30}\,\rm{G}\,\rm{cm}^3}\right)^{6/7}
	\left(\frac{M}{1.4 M_\odot}\right)^{-5/7}
	\left(\frac{\dot{M}_{\rm{in}}}{10^{3}\dot{M}_{\rm{Edd}}}\right)^{-3/7},
\end{equation}
where the constant $2 \pi \omega^{-1} k_{\rm{in}}^{3/2}$ varies in the range $1-3$
as a function of $s$. The timescale for the accretion-induced spinup of an initially
nonrotating NS to reach its equilibrium state can be roughly estimated as
\begin{equation}
	I \Delta \Omega_{\rm{NS}} \simeq \dot{M}_{\rm{in}} R_{\rm{M}}^2
	 \Omega_{\rm{s}}\left(R_{\rm{M}}\right) \Delta t,  
\end{equation}
where $I \sim M R_{\rm{NS}}^2$ is the moment of inertia of the NS, leading to
\begin{equation}\label{dt}
	\Delta t \simeq \frac{I}{\dot{M}_{\rm{in}} R_{\rm{M}}^2} = 
	1.7 \times 10^3\ \rm{yr}\, k_{\rm{in}}^{-2} \left(\frac{\mu}{10^{30}\,\rm{G}\,\rm{cm}^3}\right)^{-8/7}
	\left(\frac{M}{1.4 M_\odot}\right)^{9/7}
	\left(\frac{R_{\rm{NS}}}{10\ \rm{km}}\right)^{2}
	\left(\frac{\dot{M}_{\rm{in}}}{10^{3}\dot{M}_{\rm{Edd}}}\right)^{-3/7},
\end{equation}
within which the accretion process has negligible impact on 
the mass and radius of the NS. Correspondingly, the NS spinup rate is
\begin{equation}\label{Pd}
\frac{2\pi I \dot{P}}{P^2} \simeq \dot{M}_{\rm{in}} \sqrt{G \rm{M} R_{\rm{M}}}, \ \ \
	\dot{P} =
	9.8 \times 10^{-10}\,\rm{s}\,\rm{s^{-1}} k_{\rm{in}}^{1/2} 
	\left(\frac{\mu}{10^{30}\,\rm{G}\,\rm{cm}^3}\right)^{2/7}
	\left(\frac{M}{1.4 M_\odot}\right)^{-4/7}
	\left(\frac{R_{\rm{NS}}}{10\ \rm{km}}\right)^{2}
	\left(\frac{\dot{M}_{\rm{in}}}{10^{3}\dot{M}_{\rm{Edd}}}\right)^{6/7}
	\left(\frac{P}{1\,\rm{s}}\right)^{2}.
\end{equation}
Therefore, in the case of super-Eddington accretion, an NS can readily undergo 
rapid spinup given a high mass accretion rate, as observed in ULXP sources, 
such as the NGC 5907 ULXP with an X-ray luminosity 
$\geqslant 10^{41}\ \rm{erg} \ \rm{s}^{-1}$ \citep[]{Israel17}, provided that 
the luminosity accurately reflects the actual accretion rate\footnote{
	Among the $O(10^3)$ observed ULX candidates \citep{Walton22}, only a few
	sources exhibit pulsations, providing confirmation of their central object 
	being an NS, as shown in Table $2$ of \cite{King23}. Due to the limited number 
	of ULXPs, conducting statistical studies to investigate the general properties 
	of these sources is currently unfeasible. The observed X-ray luminosity of the
	ULXPs ranging between $10^{39}-10^{41}\,\rm{erg}\,\rm{s^{-1}}$ indicates a range of 
	the maximum NS accretion rate $10^{2}-10^{4}\dot{M}_{\rm{Edd}}$. For a strong 
	magnetic field of $\mu\gtrsim10^{31}\,\rm{G}\,\rm{cm}^3$ \citep[e.g.][]{Chashkina19}, 
	the equilibrium period is approximately $P_{\rm{eq}} \sim 0.08-0.58\,\rm{s}$,
	considering the geometrical beaming would decrease $\dot{M}_{\rm{in}}$ to further 
	raise the period to $O(1)\,\rm{s}$. Despite $P_{\rm{eq}}=O(10^{-2})\,\rm{s}$,
	the lifetime of the NS accreting system in ULX mode remains uncertain 
	\citep{Fabrika21}, thereby precluding its eventual evolution toward an equilibrium 
	state. The properties of the ULXP systems deserve further investigation.}, 
or within AGN disk environments, where the NS accretion 
rate can reach $10^5-10^6 \dot{M}_{\rm{Edd}}$ \citep[e.g.][]{Chen23}. 
Moreover, the magnetic field experiences decay throughout the NS 
accretion process \citep[e.g.][]{Mukherjee17, Konar17, Igoshev21}, 
resulting in a reduction of the equilibrium spin period $P_{\rm{eq}}$ due to an 
increased inflow-specific angular velocity at smaller $R_{\rm{M}}$. However, 
the decrease in inflow angular momentum leads to slower spinup, as indicated
by Equation (\ref{dt}) and (\ref{Pd}). Besides, through the analysis of ULXPs,
one can derive $P$ and $\dot{P}$ while estimating $\dot{M}_{\rm{in}}$ for the system,
thereby establishing a lower limit on the strength of the NS magnetic field. The properties
of the super-Eddington disk wind slightly influence the estimation of the NS magnetic field 
strength, as indicated by Equation (\ref{Pd}), where $\mu \propto k_{\rm{in}}^{-7/4}$ while 
$k_{\rm{in}}$ is not significantly relying on $s$. But the geometrical beaming originating
from the wind leads to an overestimation of $\dot{M}_{\rm{in}}$, thus $\mu$ is significantly 
underestimated, since $\mu \propto \dot{M}_{\rm{in}}^{-3}$.

\subsection{Error from radiation cooling}

\begin{figure*}
	\begin{center}
		\includegraphics[width=0.45\textwidth]{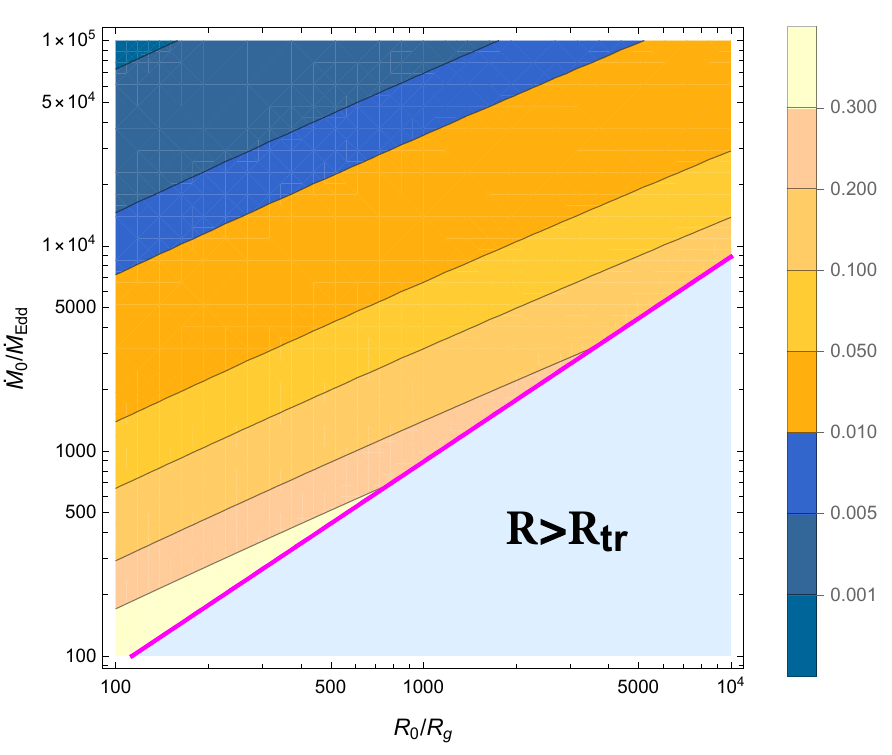}
	\end{center}
	\caption{Radiation cooling as a percentage of total disk energy,  
	$Q_{\rm{rad}}/(Q_{\rm{vis}}+Q_{\rm{rad}})$, in the $R_{\rm{o}}-
	\dot{M}_{\rm{o}}$ parameter space in the case of $s=0.5$, 
	where the energy rates are
    divided from the self-similar solutions, roughly describing the
    derivation of the solutions ignoring radiation cooling from the 
    real disk. The magenta line and the region to the tight represent the 
    case of $R_{\rm{o}} \geqslant R_{\rm{tr}}$, where the disk 
    photons can efficiently diffuse out and the disk wind would fail 
    to launch, therefore the super-Eddington disk solution is invalid 
    and the disk transitions to be geometrically thin. }	
	\label{Fig:radiation}
\end{figure*}

\begin{figure*}
	\begin{center}
		\includegraphics[width=0.45\textwidth]{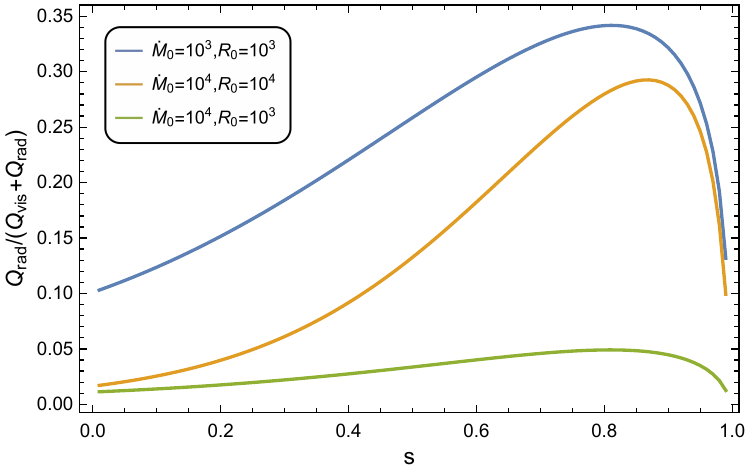}
	\end{center}
	\caption{Radiation cooling ratio 
	$Q_{\rm{rad}}/(Q_{\rm{vis}}+Q_{\rm{rad}})$ as a function 
	of $s$, where three cases are taken as examples: 
	$\dot{M}_{\rm{o}}=10^3\dot{M}_{\rm{Edd}}$ with 
	$R_{\rm{o}}=10^3R_{\rm{g}}$, $\dot{M}_{\rm{o}}=10^4\dot{M}_{\rm{Edd}}$ 
	with $R_{\rm{o}}=10^4R_{\rm{g}}$, and 
	$\dot{M}_{\rm{o}}=10^5\dot{M}_{\rm{Edd}}$ with $R_{\rm{o}}=10^4R_{\rm{g}}$.}	
	\label{Fig:radvar}
\end{figure*}

To construct a self-similar super-Eddington accretion disk, we have neglected
the radiation cooling term in Equation (\ref{energy}). In this subsection, we 
investigate whether $Q_{\rm{rad}}$ is indeed a small factor that does not 
significantly impact the disk structure. Roughly speaking, the self-similar disk 
should produce some additional energy, except its initial $Q_{\rm{vis}}$ for the 
runaway radiation. So, using Equation (\ref{Qrad}), we calculate the radiation
cooling that the self-similar disk is supposed to generate and compare it with the 
total energy of the disk encompassing this energy, i.e., $Q_{\rm{vis}}+Q_{\rm{rad}}$.
Figure \ref{Fig:radiation} and \ref{Fig:radvar} illustrate
$Q_{\rm{rad}}/(Q_{\rm{vis}}+Q_{\rm{rad}})$ as a function of $R_{\rm{o}}-\dot{M}_{\rm{o}}$
and $s$, respectively. From the figures, the ratio is consistently 
much less than unity, indicating that 
radiation cooling would not significantly impact the 
structure of the super-Eddington disk, 
except for the case of $R_{\rm{o}} \sim R_{\rm{tr}}$ 
yet $\dot{M}_{\rm{o}}/\dot{M}_{\rm{Edd}} \lesssim 1000$, 
where the disk size becomes relatively small, hindering efficient photon
advection within it. Both increasing $\dot{M}_{o}$ and decreasing $R_{\rm{o}}$ to 
achieve $R_{\rm{o}} \ll R_{\rm{tr}}$ can mitigate the impact of radiation cooling,
thereby promoting self-similarity in super-Eddington disks; two concrete disk structures
are shown in the Appendix \ref{appendix1} to illustrate this property.

The disk radiation cooling, in addition, does not exert a significant effect on the
accretion process of the NS. The magnetosphere truncation radius is directly determined by 
the disk pressure, which only exhibits a slight deviation from the self-similar value 
when considering radiation effects, as shown in Figure \ref{Fig:disk}, as $R_{\rm{M}}$.
Or, more intuitively, the inclusion of radiation cooling requires an increase of 
disk inflow mass rate to generate more energy. Given that $Q_{\rm{rad}}$ 
is much lower than $Q_{\rm{vis}}$, the alteration in the mass rate would thus be 
substantially smaller than $\dot{M}_{\rm{o}}$;
since $R_{\rm{M}} \propto \dot{M}_{\rm{o}}^{2/(7+2s)}$, 
its relative change can be neglected.

\subsection{Caveat}
In this paper, we construct the radial structure of the 
super-Eddington disk and adopt the vertical hydrostatic
equilibrium assumption. Although the resulting self-similar 
solution can conveniently provide the disk properties under given
initial conditions, they should be regarded as approximations.
In reality, the analytical solution for the disks should incorporate a 
three-dimensional or at least two-dimensional structure and discard 
the vertical force balance assumption, as the formation of disk 
winds indicates the dynamic nature of the vertical gas; only by 
constructing a stratified disk structure can we better predict the
properties of the disk wind and inflow \citep[e.g.][]{Cao22, Jiao23}.

Besides, when investigating the magnetosphere-disk interaction, the NS 
magnetic field is simplified as a dipole. However, the presence of an
accretion disk would result in the compression of the magnetosphere 
into a nondipole shape, thereby impacting the disk truncation 
\citep[e.g.][]{Aly80, Kulkarni13}. In addition,
accretion onto the NS solid surface introduces radiation feedback, which
not only induces an outward displacement of the disk inner boundary, 
as shown in Figure \ref{Fig:k}, but also brings about significant 
alterations to the disk vertical structure 
\citep[e.g.][]{Takahashi17, Takahashi18}. Furthermore, the disk wind 
would interact with the open field lines \citep[e.g.][]{Mushtukov23},
leading to reciprocal influences on their respective properties, which would
thereby alter the magnetic field structure and affect the disk cooling and
angular momentum transfer.
 
In a word, to provide a more precise depiction of the super-Eddington 
magnetized NS accretion system, it is imperative to construct a 
two-dimensional disk and explore the intricate interplay among the disk, 
the NS magnetic field, the disk wind, and the NS radiation, 
which will be studied elsewhere.

\section{Summary}
\label{section5}
In this work, we have explored the properties of a super-Eddington
magnetized NS accretion system. To comprehensively analyze the system, 
we have developed general self-similar solutions for the disk structure 
by considering variable properties of the disk wind, including its 
mass rate, specific energy, and angular momentum. Extending the disk 
toward the vicinity of the NS, we have investigated the specific interaction 
between the disk and the NS magnetosphere, finding that the entire 
accretion process exhibits similar properties to a typical system involving 
a sub-Eddington Keplerian disk, where the magnetosphere truncation radius
is approximately proportional to the Alfv\'{e}n radius, and the interaction 
boundary layer is thin. The magnetosphere truncation is slightly influenced 
by variations in the super-Eddington disk wind, unless the wind exhibits 
significant strength; conversely, it is predominantly affected by the 
advection and twisting of the NS magnetic fields. Additionally, the 
rotation of the NS and the radiation feedback from its accretion column 
are secondary factors. The resulting proportional coefficient varies 
within the range of $0.34-0.71$. Although containing simplification, 
we suggest that our model can provide a preliminary
description of the accretion system, thus serving as a foundation for 
constructing a more accurate system structure or for studying the evolution
of NSs under super-Eddington disk accretion.

\begin{acknowledgements}
	
This work was supported by the National SKA Program of China 
(grant No. 2020SKA0120302), and the National Natural Science 
Foundation of China (grant No. 12393812).
		
\end{acknowledgements}

\begin{appendix}
\section{Numerical Solutions of Super-Eddington Accretion Disk }
\label{appendix1}
\begin{figure*}
	\begin{center}
		\includegraphics[width=1\textwidth]{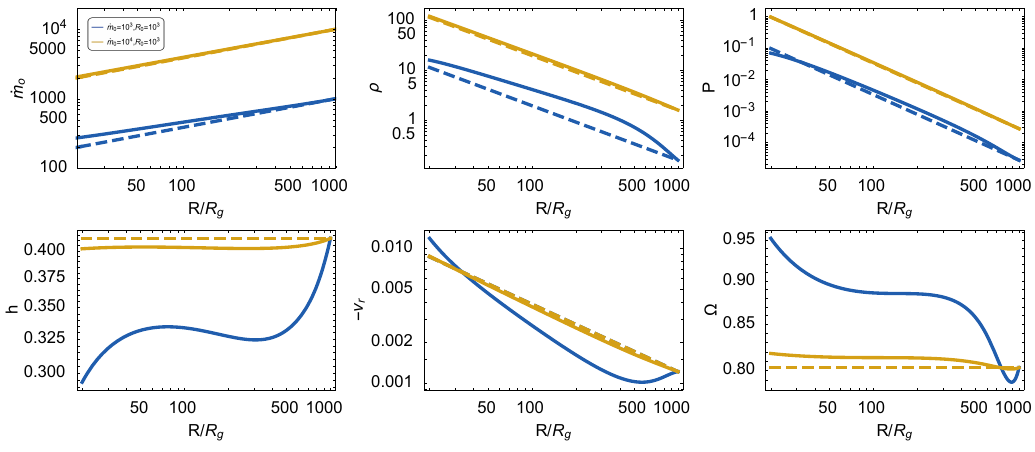}
	\end{center}
	\caption{The deviation between numerical and analytic solutions of 
	super-Eddington accretion disks, taking a self-similar solution 
	as the boundary conditions. In each panel, the blue lines show the
    case of $\dot{M}_{\rm{o}}=10^3\dot{M}_{\rm{Edd}}, R_{\rm{o}}=10^3
    R_{\rm{g}}$, and the golden lines show the case of 
    $\dot{M}_{\rm{o}}=10^4\dot{M}_{\rm{Edd}}, R_{\rm{o}}=10^3 R_{\rm{g}}$;
    the dashed lines are the self-similar solutions ignoring radiation
    cooling, and the solid lines are the structures derived from numerically 
    solving the disk equations directly.
    All the disk parameters are dimensionless, with $\dot{m}_{\rm{o}}, \rho,
    P, h, v_{\rm{r}}$, and $\Omega$ as units of $\dot{M}_{\rm{Edd}}, 
    \dot{M}_{\rm{Edd}}/\sqrt{G M R_{\rm{g}}^3}, \dot{M}_{\rm{Edd}} 
    \sqrt{G M / R_{\rm{g}}^5}, R_{\rm{g}}, \sqrt{G M / R_{\rm{g}}}$,
    and $\sqrt{G M / R_{\rm{g}}^3}$, respectively. }	
	\label{Fig:disk}
\end{figure*}

To directly demonstrate the impact of neglecting the radiation cooling 
term in Equation (\ref{energy}) on disk structures, we employ the disk 
self-similar solutions at the outer radius as boundary conditions and 
numerically solve differential Equations (\ref{EqM})-(\ref{energy}) 
inward to determine the resultant disk structure. To 
specify the property of the disk wind, we adopt $s=\lambda(H/R)$ with
$\lambda=1$, and the obtained value of $s$ is $0.412$. Two
examples, $\dot{M}_{\rm{o}}=10^3\dot{M}_{\rm{Edd}}, R_{\rm{o}}=10^3
R_{\rm{g}}$ representing scenarios where $R_{\rm{o}} \sim R_{\rm{tr}}$, and
$\dot{M}_{\rm{o}}=10^4\dot{M}_{\rm{Edd}}, R_{\rm{o}}=10^3
R_{\rm{g}}$ representing scenarios where $R_{\rm{o}} \ll R_{\rm{tr}}$, are
shown in Figure \ref{Fig:disk}. From the figure, the numerical solution 
exhibits minimal deviation from the self-similar one when
$R_{\rm{o}} \ll R_{\rm{tr}}$, as the contribution of the radiation cooling 
is negligible, which is also illustrated in Figure \ref{Fig:radvar}. But
when $R_{\rm{o}} \sim R_{\rm{tr}}$, the proportion of the radiation cooling is 
increased, resulting in a slight deviation of 
$\dot{m}_{\rm{o}}, \rho$, and $P$. Additionally, due to the crucial 
role of boundary conditions in solving differential equations, 
selecting the self-similar values for inwardly numerical calculations
would lead to an accumulative deviation of
$h, v_{\rm{r}}$, and $\Omega$. On the whole, the analytic self-similar 
solution exhibits a reasonable level of validity, while the numerical 
solution of a super-Eddington accretion disk deserves further investigation.

\end{appendix}

\end{document}